# Rate Splitting for 6G Optical Wireless Networks

Khulood D. Alazwary[1], Ahmad Adnan Qidan[2], T. E. H. El-Gorashi[2] and Jaafar M. H. Elmirghani[2]
[1]School of Electronic and Electrical Engineering, University of Leeds, LS2 9JT, United Kingdom
[2]Department of Engineering, Faculty of Natural, Mathematical & Engineering Sciences Kings College London
*elkal@leeds.ac.uk, ahmad.qidan@kcl.ac.uk, taisir.elgorashi@kcl.ac.uk, jaafar.elmirghani@kcl.ac.uk*

**ABSTRACT**
This paper evaluates the performance of rate splitting (RS), a robust interference management scheme, in an optical wireless communication (OWC) network that uses infrared lasers referred to as vertical-cavity surface-emitting lasers (VCSELs) as optical transmitters. In 6G OWC, providing high spectral and energy efficiency requires advanced multiple access schemes that can serve multiple users simultaneously in a non-orthogonal fashion. In this context, RS has the potential to manage multi-user interference at high data rates compared to orthogonal transmission schemes. Simulation results show the high performance of RS compared to baseline approaches.

## 1. INTRODUCTION

Current radio frequency (RF) networks face several challenges in terms of lack of resources, energy efficiency and security. Therefore, complementary technologies are required to support the high demands expected in the next generation (6G) of wireless communications. Optical wireless communication (OWC) has received massive interest as a strong candidate in 6G networks. OWC offers license-free bandwidth and improved security, and usually optical access points (APs) provide high energy efficiency [1]–[8]. OWC networks can provide aggregate data rates in a range of several gigabits per second (Gbps) using conventional light emitting diodes (LEDs). Further investigation is needed to tackle OWC challenges such as light blockage, the confined converge area of optical APs, and the low modulation speed of LED light sources [4]–[13]. In this context, different techniques has been introduced to improve the capacity of optical communication links such as beam power, beam angle, and beam delay adaptations [3], [14]–[21]. Moreover, using VCSELs as optical transmitters is proposed since they have high modulation speeds compared to LEDs, and can provide up to terabits per second (Tbps) aggregate data rates [22]–[25]. The use of VCSELs as optical transmitters is envisaged as a key enabler for the next generation of OWC networks [26], [27]. In particular, VCSELs are expected to provide ubiquitous connectivity and high spectral and energy efficiency under eye safety constraints [28].

Interference management (IM) is crucial in multiple-input and multiple-output (MIMO) OWC scenarios where multiple users must be served simultaneously. Accordingly, some orthogonal transmission schemes were implemented to avoid interference such as time division multiple access (TDMA) [29], orthogonal frequency division multiple access (OFDMA) [30], optical code division multiple access (OCDMA) [31], wavelength division multiple access (WDMA) [25] and space division multiple access (SDMA) [32]. Contrary to orthogonal transmission schemes, non-orthogonal multiple access (NOMA) [33] has been proposed for OWC to manage multi-user interference through the power domain where power allocation is performed considering the channel quality of each user. Nevertheless, such transmission schemes suffer from the inefficient use of available resources where the number of users is relatively large, and can be even more challenging in practice. To address such a problem, rate splitting (RS) is a known solution that utilises spatial domain and power domain to enhance the system performance in multi-user multiple-input single-output (MU-MISO) systems. Basically, each users' messages in RS is split in into common and private parts. The common parts of all users in the network are jointly encoded into one common stream, while the private parts are encoded independently. At the user side, the common stream is first decoded by treating all the private streams as noise and then the user's private stream can be recovered using Successive Interference cancellation (SIC). A study in [34] shows that RS in RF provides better energy and spectrum efficiencies than NOMA. In [35], the spectral and energy efficiency trade-off are studied in MISO broadcast channel scenarios using RS. In OWC, RS has the ability to provide high performance in overloaded networks by using parallel RS schemes working simultaneously among multiple groups of users [22], [36]. Furthermore, optimum RS for OWC was proposed in [37] where a rate maximization optimization problem was formulated to allocate the power optimally among the common and private messages of RS.

Motivated by the benefits of RS in overcoming the limitations of conventional IM schemes, in this paper, we study and validate the performance gain of RS in laser-based OWC networks.

The rest of this paper is organized as follows: Section 2 describes the system configuration, including the room layout, VCSEL and receiver configuration. RS is discussed in Section 3. The simulation results are shown and discussed in Section 4 and conclusions are stated in Section 5.

## 2. SYSTEM MODEL:

We consider a downlink OWC network as shown in Figure 1. The system has $L$ optical APs. Each AP is composed of multiple VCSELs, distributed on the ceiling to serve $K$ users. Each user is equipped with angle diversity receiver (ADR) as in [25], which has $M$ photodiodes. Each photodiode points to different directions allowing users to connect to most of the available optical APs. In the considered indoor environment, the users are randomly distributed over the communication floor which is 0.85 m above the ground. The transmitted signal can be written in a vector form as

$$\mathbf{x} = [x_1 \quad x_2 \quad \ldots \quad x_L]^T \in \mathbb{R}_+^{L \times 1}, \quad (1)$$

where $x_l$ is the signal transmitted by optical transmitter $l$. At this point, the received signal of user $k$ regardless of the multiple access scheme used is given by

$$y^{[k]} = \mathbf{h}^{[k]}\mathbf{x} + z^{[k]}, \quad (2)$$

where $\mathbf{h}^{[k]}$ is the channel vector between user $k$ and the $L$ optical transmitters and $z^{[k]}$ is real valued additive white Gaussian noise (AWGN) with zero mean and variance $\sigma_k^2$, which represents the sum of thermal noise and shot noise. Note that all the optical APs are connected through a central unit (CU), which has important information to control the network. Table 1 shows all the simulation parameters.

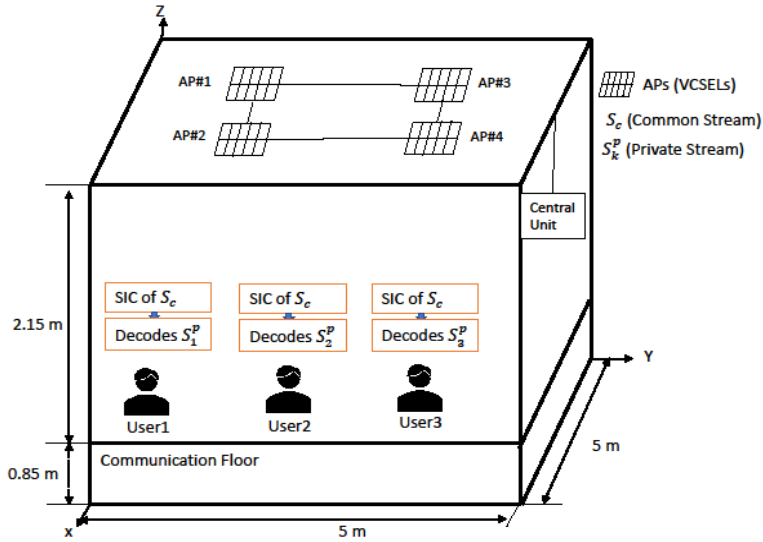

Figure 1: Room Configuration

We consider the VCSEL as an optical transmitter, whose power distribution can be determined in accordance to the beam waist $W_0$, the wavelength $\lambda$ and the distance $d$ from the ceiling to the communication floor. In this regards, the beam radius of the VCSEL at user $k$ located on the communication floor at distance $d$ is given by

$$W_d = W_0 \left(1 + \left(\frac{d}{d_{Ra}}\right)^2\right)^{\frac{1}{2}}, \quad (3)$$

where $d_{Ra}$ is the Rayleigh distance that can be expressed as

$$d_{Ra} = \frac{\pi W_0^2 n}{\lambda}, \quad (4)$$

where $n$ is the refractive index and $n = 1$ for air. Accordingly, the spatial intensity profile of each VCSEL $l$ over the transverse plane at distance $d$ is given by

$$I_l(r,d) = \frac{2P_{t,l}}{\pi W_d^2} \exp\left(-\frac{2r^2}{W_d^2}\right), \quad (5)$$

where $P_{t,l}$ is the optical power and $r$ is the radial distance from the centre of the beam spot and the distance $d$. At this point, the received power at photodiode $m$ at user $k$ from the VCSEL $l$ can be calculated as

$$P_{m,l} = \int_0^{r_m} I(r,d) 2\pi r \, dr = P_{t,l}\left[1 - \exp\left(\frac{-2r_m^2}{W_d^2}\right)\right] \quad (6)$$

where $r_m$ is the radius of photodiode $m$. Assuming that all the photodiodes of the ADR have the same size and detection area, the detection area of photodiode $m$ is determined by $A_m = \frac{A_{rec}}{M}$, where $A_{rec}$ is the whole detection area of the ADR. More details on the received power calculations of VCSEL can be found in [28].

**Table 1**. Simulation Parameters

| VCSEL parameter | Value |
|---|---|
| Laser Beam waist, W0 | 5-30 μm |
| Laser Bandwidth | 5 GHz |
| Laser Wavelength | 850 nm |
| Receiver Responsivity | 0.4 A/W |
| Receiver FOV | 45 deg |
| Area of the photodetector | 20 mm2 |
| Receiver noise current spectral density | 4.47 pA/√Hz |
| Receiver Gain filter | 1.0 |
| Room ( Width × Length × Height ) (x, y, z) | 5 m × 5 m × 3 m |

## 3. RATE SPLITTING:

RS divides each user's message into two sub-messages, common and private messages, each message is transmitted at a different transmit power. The common messages of all $K$ users are encoded into one super common message denoted by $s_c$, superimposed on top of the encoded private messages of the $K$ users given by $\mathbf{s_p} = [s_p^{[1]} \; s_p^{[2]} \; ... \; s_p^{[K]}]$. Thus, the transmit signal includes one common message and $K$ private messages can be expressed as:

$$\mathbf{x} = \sqrt{P_c}\mathbf{w}_c s_c + \sum_{k=1}^{K} \sqrt{P_p}\mathbf{w}_p^{[k]} s_p^{[k]} \qquad (7)$$

where $\mathbf{w}_c$ is the unit-norm precoding vector of the common message and $\mathbf{w_p} = [\mathbf{w}_p^{[1]} \mathbf{w}_p^{[2]} ... \mathbf{w}_p^{[K]}]$ is the precoding vectors of all $K$ private messages. For simplicity, a fixed power allocation approach is applied among the common and private messages, where $P_c$ and $P_s$ are the powers of the common and private messages, respectively. Taking into consideration the total power $P_T$, the power of the single private message is obtained as $P_p = \frac{P_T \alpha}{K}$, where $\alpha \in \{0,1\}$, while the residual power is allocated to the common message $P_c = P_T (1 - \alpha)$. The received signal of user $k$ can be expressed as

$$y^{[k]} = \sqrt{P_c}\mathbf{h}^{[k]}\mathbf{w}_c s_c + \sqrt{P_p}\mathbf{h}^{[k]}\mathbf{w}_p^{[k]} s_p^{[k]}$$
$$+ \underbrace{\sum_{k' \neq k}^{K} \sqrt{P_p}\mathbf{h}^{[k]}\mathbf{w}_p^{[k']} s_p^{[k']}}_{\text{multi-user interference}} + z^{[k]} \qquad (8)$$

The decoding procedure at each user $k$ starts by decoding the common message while treating all the private messages as noise then using SIC technique to decode its private message. At this point, the SINR equations of the common and private messages can be derived as

$$\gamma_c^{[k]} = \frac{P_c \left|\mathbf{h}^{[k]H}\mathbf{w}_c\right|^2}{\sum_{k=1}^{K} P_p \left|\mathbf{h}^{[k]H}\mathbf{w}_p^{[k]}\right|^2 + \sigma_z^2} \qquad (9)$$

$$\gamma_p^{[k]} = \frac{P_p \left|\mathbf{h}^{[k]H}\mathbf{w}_p^{[k]}\right|^2}{\sum_{k' \neq k} P_p \left|\mathbf{h}^{[k]H}\mathbf{w}_p^{[k']}\right|^2 + \sigma_z^2} \qquad (10)$$

respectively. The achievable rate of the common message is $R_c = \log_2 \left(1 + \min_k \{\gamma_c^{[k]}\}\right)$, where $\min_k \{\gamma_k^c\}$ ensures that each user $k$ can successfully decode the common message, while the sum rate of the private messages of all the users is $R_p = \sum_{k=1}^{K} \log_2 \left(1 + \gamma_p^{[k]}\right)$. Therefore, the sum rate of the network considering RS is $R_{RS} = R_c + R_p$.

## 4. SIMULATION SETUP AND RESULTS:

In this section, we evaluate the performance of RS in an OWC indoor environment with 5 m × 5 m × 3 m dimensions. Our network is composed of $L = 4$ optical APs deployed on the ceiling, each AP is an array of

VCSELs to provide uniform coverage to $K = 10$ users distributed on the communication floor. The rest of the simulation parameters are illustrated in Table I.

In Fig. 2, the sum rate of RS is shown against a range of SNR values from 5 dB to 25 dB. It can be seen that the optimum RS we proposed in [37] provides higher sum rates at all the SNR values compared to conventional RS and orthogonal transmission schemes due to the fact that the power is allocated to common and private messages of users with the aim of maximizing the sum rate of the network. On the other hand, conventional RS allocates the power uniformly among the users, and OMA avoids the interference at the cost of low spectral efficiency.

Fig. 3 compares the sum rate of the considered schemes versus various values of the VCSEL beam waist. It is observed that the optimum RS results in enhancing the sum rate of the network considerably compared with the performance of conventional RS and OMA schemes. Moreover, the sum rate increases with the beam waist $W_0$ regardless of the scheme considered. The reason is that an increase in the beam waist results in a high received power and less beam divergence. However, eye safety recommendations must be considered.

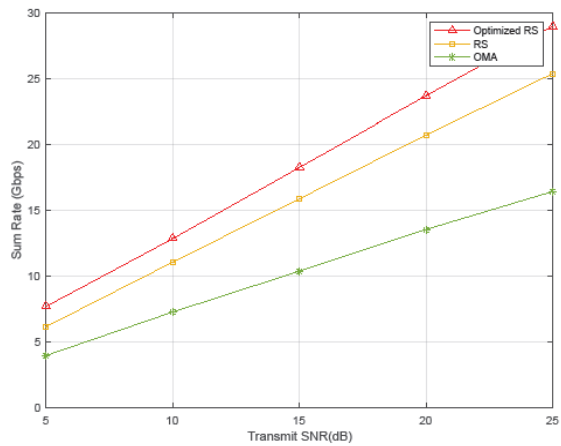

Figure 2: Sum rates versus SNR

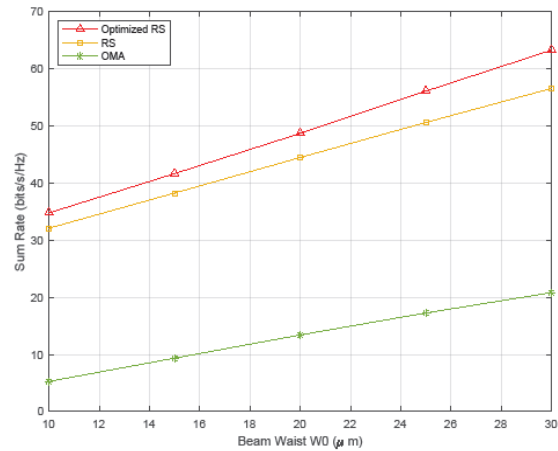

Figure 3: Sum rates versus $W_0$

## 5. CONCLUSIONS

We have studied the performance of RS in a laser-based OWC system. The simulation results showed that RS with an optimum power allocation approach achieves 13% higher data rates compared to conventional RS, which considers uniform power allocation, and 76% higher than orthogonal multiple access. Moreover, the impact of the VCSEL beam waist $W_0$ on the sum rate is shown. For future work, complex optimization problems will be formulated to further maximize the sum rate of RS under eye safety regulations.

## ACKNOWLEDGEMENTS

The authors would like to acknowledge funding from the Engineering and Physical Sciences Research Council (EPSRC) INTERNET (EP/H040536/1), STAR (EP/K016873/1) and TOWS (EP/S016570/1) projects. The first author would like to thank King Abdulaziz University in the Kingdom of Saudi Arabia for funding her PhD scholarship. All data are provided in full in the results section of this paper.

## REFERENCES

[1] B. Ai, A. F. Molisch, M. Rupp, and Z. D. Zhong, "5G key technologies for smart railways," Proc. IEEE, vol. 108, no. 6, pp. 856–893, 2020, doi: 10.1109/JPROC.2020.2988595.

[2] A. A. Qidan, M. Morales Cespedes, and A. Garcia Armada, "Load balancing in hybrid VLC and RF networks based on blind interference alignment," IEEE Access, vol. 8, pp. 72 512–72 527, 2020

[3] A. A. Qidan, T. El-Gorashi, and J. M. H. Elmirghani, "Cooperative Artificial Neural Networks for Rate-Maximization in Optical Wireless Networks," in ICC 2023 - IEEE International Conference on Communications, 2023.

[4] A. T. Hussein and J. M. H. Elmirghani, "10 Gbps Mobile Visible Light Communication System Employing Angle Diversity, Imaging Receivers, and Relay Nodes," J. Opt. Commun. Netw., vol. 7, no. 8, p. 718, 2015, doi: 10.1364/jocn.7.000718.

[5] A. T. Hussein and J. M. H. Elmirghani, "Mobile Multi-Gigabit Visible Light Communication System in Realistic Indoor Environment," J. Light. Technol., vol. 33, no. 15, pp. 3293–3307, 2015, doi: 10.1109/JLT.2015.2439051.

[6] M. T. Alresheedi and J. M. H. Elmirghani, "Hologram selection in realistic indoor optical wireless systems with angle diversity receivers," J. Opt. Commun. Netw., 2015, doi: 10.1364/JOCN.7.000797.

[7] A. Adnan-Qidan, M. Morales-Cespedes, A. Garcia-Armada, and J. M. H. Elmirghani, "User-centric cell formation for blind interference alignment in optical wireless networks," in ICC 2021 - IEEE International Conference on Communications, 2021, pp. 1–7.

[8] A. A. Qidan, M. Morales Cespedes, A. Garcia Armada, and J. M. Elmirghani, "Resource allocation in user-centric optical wireless

cellular networks based on blind interference alignment," Journal of Lightwave Technology, pp. 1–1, 2021

[9] O. Z. Alsulami, M. T. Alresheedi, and J. M. H. Elmirghani, "Optical Wireless Cabin Communication System," 2019 IEEE Conf. Stand. Commun. Networking, CSCN 2019, no. June, 2019, doi: 10.1109/CSCN.2019.8931345.

[10] O. Z. Alsulami, M. T. Alresheedi, and J. M. H. Elmirghani, "Transmitter diversity with beam steering," Int. Conf. Transparent Opt. Networks, vol. 2019-July, pp. 1–5, 2019, doi: 10.1109/ICTON.2019.8840147.

[11] A. S. Elgamal, O. Z. Aletri, A. A. Qidan, T. E. H. El-Gorashi and J. M. H. Elmirghani, "Reinforcement Learning for Resource Allocation in Steerable Laser-Based Optical Wireless Systems," in 2021 IEEE Canadian Conference on Electrical and Computer Engineering (CCECE), 2021.

[12] O. Z. Alsulami, M. O. I. Musa, M. T. Alresheedi, and J. M. H. Elmirghani, "Visible light optical data centre links," Int. Conf. Transparent Opt. Networks, vol. 2019-July, pp. 1–5, 2019, doi: 10.1109/ICTON.2019.8840517.

[13] S. O. M. Saeed, S. Hamid Mohamed, O. Z. Alsulami, M. T. Alresheedi, and J. M. H. Elmirghani, "Optimized resource allocation in multi-user WDM VLC systems," Int. Conf. Transparent Opt. Networks, vol. 2019-July, pp. 1–5, 2019, doi: 10.1109/ICTON.2019.8840439.

[14] F. E. Alsaadi and J. M. H. Elmirghani, "Adaptive mobile line strip multibeam MC-CDMA optical wireless system employing imaging detection in a real indoor environment," IEEE J. Sel. Areas Commun., vol. 27, no. 9, pp. 1663–1675, 2009, doi: 10.1109/JSAC.2009.091216.

[15] F. E. A. and J. M. H. Elmirghani, "High-speed spot diffusing mobile optical wireless system employing beam angle and power adaptation and imaging receivers," J. Light. Technol., vol. 28, no. 6, pp. 2191–2206, 2010.

[16] F. E. Alsaadi and J. M. H. Elmirghani, "Mobile multigigabit indoor optical wireless systems employing multibeam power adaptation and imaging diversity receivers," J. Opt. Commun. Netw., vol. 3, no. 1, pp. 27–39, 2011, doi: 10.1364/JOCN.3.000027.

[17] F. E. Alsaadi and J. M. H. Elmirghani, "Performance evaluation of 2.5 Gbit/s and 5 Gbit/s optical wireless systems employing a two dimensional adaptive beam clustering method and imaging diversity detection," IEEE J. Sel. Areas Commun., vol. 27, no. 8, pp. 1507–1519, 2009, doi: 10.1109/JSAC.2009.091020.

[18] F. E. Alsaadi, M. Nikkar, and J. M. H. Elmirghani, "Adaptive mobile optical wireless systems employing a beam clustering method, diversity detection, and relay nodes," IEEE Trans. Commun., vol. 58, no. 3, pp. 869–879, 2010, doi: 10.1109/TCOMM.2010.03.080361.

[19] M. T. Alresheedi and J. M. H. Elmirghani, "Performance Evaluation of 5 Gbit/s and 10 Gbit/s Mobile Optical Wireless Systems Employing Beam Angle and Power Adaptation with Diversity Receivers," IEEE J. Sel. Areas Commun., vol. 29, no. 6, pp. 1328–1340, 2011, doi: 10.1109/JSAC.2011.110620.

[20] M. T. Alresheedi and J. M. H. Elmirghani, "10 Gb/s indoor optical wireless systems employing beam delay, power, and angle adaptation methods with imaging detection," J. Light. Technol., vol. 30, no. 12, pp. 1843–1856, 2012, doi: 10.1109/JLT.2012.2190970.

[21] A. S. Elgamal, O. Z. Alsulami, A. A. Qidan, T. E. H. El-Gorashi, and J. M. H. Elmirghani, "Q-learning algorithm for resource allocation in WDMA-based optical wireless communication networks," 2021 6th Int. Conf. Smart Sustain. Technol. Split. 2021, 2021, doi: 10.23919/SpliTech52315.2021.9566383.

[22] K. Alazwary, A. A. Qidan, T. El-Gorashi, and J. M. H. Elmirghani, "Rate splitting in VCSEL-based optical wireless networks," 2021 6th Int. Conf. Smart Sustain. Technol. Split. 2021, pp. 1–5, 2021, doi: 10.23919/SpliTech52315.2021.9566354.

[23] A. A. Qidan, M. Morales-Cespedes, T. El-Gorashi, and J. M. H. Elmirghani, "Resource Allocation in Laser-based Optical Wireless Cellular Networks," 2021 IEEE Glob. Commun. Conf. GLOBECOM 2021 - Proc., 2021, doi: 10.1109/GLOBECOM46510.2021.9685357.

[24] W. Z. Ncube, A. A. Qidan, T. El-Gorashi, and M. H. Jaafar Elmirghani, "On the energy efficiency of Laser-based Optical Wireless Networks," Proc. 2022 IEEE Int. Conf. Netw. Softwarization Netw. Softwarization Coming Age New Challenges Oppor. NetSoft 2022, pp. 7–12, 2022, doi: 10.1109/NetSoft54395.2022.9844102.

[25] O. Z. Alsulami et al., "Optimum resource allocation in optical wireless systems with energy-efficient fog and cloud architectures," Philos. Trans. R. Soc. A Math. Phys. Eng. Sci., vol. 378, no. 2169, Apr. 2020, doi: 10.1098/rsta.2019.0188.

[26] A. Qidan, T. El-Gorashi, and J. Elmirghani, "Towards Terabit LiFi Networking," no. Photoptics, pp. 203–212, 2022, doi: 10.5220/0010955000003121.

[27] A. A. Qidan, T. El-Gorashi, and J. M. H. Elmirghani, "Artificial Neural Network for Resource Allocation in Laser-based Optical wireless Networks," IEEE Int. Conf. Commun., vol. 2022-May, pp. 3009–3015, 2022, doi: 10.1109/ICC45855.2022.9839259.

[28] M. D. Soltani et al., "Safety Analysis for Laser-based Optical Wireless Communications: A Tutorial," pp. 1–54, 2021, [Online]. Available: http://arxiv.org/abs/2102.08707.

[29] A. M. Abdelhady, O. Amin, A. Chaaban, B. Shihada, and M. S. Alouini, "Downlink Resource Allocation for Dynamic TDMA-Based VLC Systems," IEEE Trans. Wirel. Commun., vol. 18, no. 1, pp. 108–120, 2019, doi: 10.1109/TWC.2018.2877629.

[30] A. Adnan-Qidan, M. Morales-Cespedes, and A. G. Armada, "User-centric blind interference alignment design for visible light communications," IEEE Access, vol. 7, pp. 21 220–21 234, 2019.

[31] J. a. Salehi, "Code division multiple-access techniques in optical fiber networks-Part I: Fundamental principles," IEEE Trans. Commun, vol. 31, no. 8, pp. 834–842, 1989.

[32] Z. Chen and H. Haas, "Space division multiple access in visible light communications," IEEE Int. Conf. Commun., vol. 2015-Septe, no. 1, pp. 5115–5119, 2015, doi: 10.1109/ICC.2015.7249135.

[33] M. K. Aljohani et al., "NOMA visible light communication system with angle diversity receivers," in International Conference on Transparent Optical Networks, Jul. 2020, vol. 2020-July, doi: 10.1109/ICTON51198.2020.9203212.

[34] Y. Mao, B. Clerckx, and V. O. K. Li, "Rate-Splitting for Multi-Antenna Non-Orthogonal Unicast and Multicast Transmission: Spectral and Energy Efficiency Analysis," IEEE Trans. Commun., vol. 67, no. 12, pp. 8754–8770, 2019, doi: 10.1109/TCOMM.2019.2943168.

[35] G. Zhou, Y. Mao, and B. Clerckx, "Rate-Splitting Multiple Access for Multi-Antenna Downlink Communication Systems: Spectral and Energy Efficiency Tradeoff," IEEE Trans. Wirel. Commun., vol. 21, no. 7, pp. 4816–4828, 2022, doi: 10.1109/TWC.2021.3133433.

[36] A. A. Qidan, K. Alazwary, T. El-Gorashi, M. Safari, H. Haas, R. V. Penty, I. H. White, and J. M. H. Elmirghani., "Multi-User Rate Splitting in Optical Wireless Networks," pp. 1–37, 2022, [Online]. Available: http://arxiv.org/abs/2207.11458.

[37] K. Alazwary, A. A. Qidan, T. El-Gorashi, and J. M. H. Elmirghani, "Optimizing Rate Splitting in Laser-based Optical Wireless Networks," 2022 IEEE Int. Conf. Commun. Work. ICC Work. 2022, pp. 463–468, 2022, doi: 10.1109/ICCWorkshops53468.2022.9814503.